\documentclass[conference]{IEEEtran}
\IEEEoverridecommandlockouts
\usepackage{cite}
\usepackage{amsmath,amssymb,amsfonts}
\usepackage{algorithmic}
\usepackage{graphicx}
\usepackage{textcomp}
\usepackage{xcolor}
\usepackage{comment} % This is used to comment sections
\usepackage{adjustbox} % This is used to adjust table width if it is outside
\usepackage{tabularx} % for flexible table columns
\usepackage{multirow} % for multirow cells in tables
\usepackage{url} % for urls

\usepackage[linesnumbered,ruled]{algorithm2e}

\usepackage[caption=false]{subfig}

\def\BibTeX{{\rm B\kern-.05em{\sc i\kern-.025em b}\kern-.08em
    T\kern-.1667em\lower.7ex\hbox{E}\kern-.125emX}}
\begin{document}
\SetKwComment{Comment}{\footnotesize/* }{\footnotesize */}

\title{Monitoring the evolution of antisemitic discourse on extremist social media using BERT\thanks{\textbf{Acknowledgements:} Research supported by American University's Signature Research Initiative program. We also thank Matan Schwartz for his initial help with this project.}\\
}

 \author{\IEEEauthorblockN{Raza Ul Mustafa, Nathalie Japkowicz}
  \IEEEauthorblockA{\textit{American University, 4400 Massachusetts Ave NW, Washington, DC 20016, USA} \\
 \{rmustafa,japkowic\}@american.edu}
 }

\maketitle

\begin{abstract}
Racism and intolerance on social media contribute to a toxic online environment which may spill offline to foster hatred, and eventually lead to physical violence. 
That is the case with online antisemitism, the specific category of hatred considered in this study.  Tracking antisemitic themes and their associated terminology over time in online discussions could help monitor the sentiments of their participants and their evolution, and possibly offer avenues for intervention that may prevent the escalation of hatred.
Due to the large volume and constant evolution of online traffic, monitoring conversations manually is impractical. Instead, we propose an automated method that extracts antisemitic themes and terminology from extremist social media over time and captures their evolution. Since supervised learning would be too limited for such a task, we created an \textit{unsupervised online machine learning} approach that uses large language models to assess the contextual similarity of posts. The method clusters similar posts together, dividing, and creating additional clusters over time when sub-themes emerge from existing ones or new themes appear. The antisemitic terminology used within each theme is extracted from the posts in each cluster. Our experiments show that our methodology outperforms existing baselines and demonstrates the kind of themes and sub-themes it discovers within antisemitic discourse along with their associated terminology. We believe that our approach will be useful for monitoring the evolution of all kinds of hatred beyond antisemitism on social platforms.
\end{abstract}


\section{Introduction}
Social\footnote{Warning: This paper contains examples of racist and antisemitic statements that may be disturbing to the reader.} media has played a crucial role in simplifying the human process of joining communities of like-minded individuals. People use these platforms to engage with others by way of expressing their opinions and commenting on those of others. While technology has had a positive impact on many people's lives, it has also brought serious negative consequences onto society, including an acceleration of the spread of hatred.

In the past few years, %Though many 
a large number of studies have been devoted to automatic hate speech detection \cite{Schmidt2017ASO, Fortuna2018ASO, Poletto2020ResourcesAB, Jahan2021ASR}. The more difficult task of understanding the nature of online discussions automatically has been considered, though only rarely. On those few occasions, the task was considered in a static fashion, ignoring the fact that online discussions evolve quickly. Though these attempts constitute a good start, they are not sufficient since %Yet, such a goal is important for 
%both governmental and non-governmental 
organizations %alike 
that seek to promote social peace need to be informed of changes in discussion themes as they occur rather than on, say, a monthly basis, when the software is run on a new batch of posts.
%
%, since getting a better handle on the semantic themes expressed in these discussions could help combat online hate speech through targeted education, intervention, and possibly other actions.
The purpose of our work is to explore the ability of machine learning (ML) and natural language processing (NLP) tools to extract a continuously evolving map of the semantic themes represented in social media posts, automatically.  To provide continuity and avoid catastrophic forgetting, it is of utmost importance that past knowledge be preserved and only  updated when necessary.
%with the arrival of new data, while new information is added to the existing knowledge base.  

We develop a method to continuously monitor antisemitic themes from discussions taking place on extremist social media. Many antisemitic themes have emerged throughout the years, making antisemitism an unfortunately good testbed for our approach. Indeed, Jews have been victims of discrimination for centuries, sometimes accused of having too much power, controlling the world economy, or being disloyal. Discussions and conspiracy theories about the Holocaust abound. %, along with suggestions that it was invented to benefit the Jews. 
Other historical tropes include: ``Jews killed Jesus" and ``Jews are greedy". These conspiracy theories evolve continuously developing new themes, such as linking Jews to the COVID-19 pandemic \cite{barna2021exploration}.
The purpose of our endeavor is to stay ahead of the "mood" in extremist social media communities where antisemitic posts are frequent, so as to be aware of emerging new themes or recurring old ones in the same or altered form.  To be practical, given the exploratory and continuous nature of the task, the method we devised is unsupervised and works in a continual and, thus, stable fashion using state-of-the-art ML and NLP techniques such as Large Language Models (LLMs). Though antisemitism was used in this work as a case study, the methods we propose will apply to any other kinds of hatred, such as Islamophobia, anti-Black or anti-Asian racism, anti-LGBTQ+ discourse, etc. 

%In our work in the context of antisemitic hate speech, the 
Our proposed approach %is an online approach that 
begins by considering posts collected in a first time-window and groups together semantically related themes (i.e., concepts, topics). %, and extracting antisemitic terminology (common features) from them. 
It then proceeds with the next time-window, adding new posts to the previously discovered themes and discovering new themes in the data if the discourse they represent stands semantically apart from %that of historical tropes or 
the themes already present in memory %{\color{red} These new themes are considered stand-alone themes if they differ significantly from themes already present NOT SURE THAT'S TRUE}, 
and sub-themes, if they share enough commonality with existing ones. The system then considers the posts collected in the next time-window and repeats the same steps over and over. 
\begin{comment}

In addition to proposing a practical approach,
%Note: We use the terms tropes and concepts interchangeably throughout the paper since both convey a similar meaning in this research. If we summarize 
this study answers %three 
the following research question: 
\begin{description}
%    \item How can the transformer architecture of large language models be used for automatic antisemitic tropes or themes discovery?
%    \item Can large language models easily extract the antisemitic terminology floated on extremist social media platforms from the themes they discovered?
 %   \item Are the themes and their related terminology extracted from the posts coherent and useful? %What is the distributional semantics of antisemitic terminology and tropes?
\item[RQ:] How can the transformer architecture of large language models be used for constructing an \textbf{evolving map of antisemitic themes} automatically, and extracting \textbf{new hate-laden terminology} from it?
%\item Are the themes and their related terminology extracted from the posts coherent and useful?
\end{description} 
\end{comment}

The remainder of the paper is organized as follows. In Section \ref{tab:background}, we discuss the philosophical basis for our work along with related studies from the fields of Computer Science, Psychology and Business. We then review ML and NLP techniques previously used for combating online hate speech and antisemitism in Section~\ref{tab:literature}. %Section \ref{tab:approach} introduces our dataset collection procedure. 
Section \ref{tab:approach} describes our methodology. %approach for automatic antisemitic tropes discovery and antisemitic terms extraction using contextual embedding. Next, we compare our approach with a few other techniques in Section \ref{tab:comparison_with_other_approaches}. 
We present and discuss our results in Section \ref{tab:results_and_discussion} and conclude our paper in Section \ref{tab:conclusion}.

\section{General Background} \label{tab:background}
In this section, we consider related work from four different disciplines: Philosophy, Computer Science, Psychology, Business.
\subsubsection*{Philosophical Concept Theory}
Our work is based on Philosophical Concept  Theory \cite{Allen2003StanfordEO} used extensively in the fields of Psychology and Cognitive Science. In a nutshell, we take the view that concepts are representations of the mind and that complex concepts are composed of sub-concepts. Furthermore, we subscribe to the Prototype Theory rooted in Wittgenstein \cite{Wittgenstein1958PhilosophicalI} and Rosch \cite{Rosch1978PrinciplesOC}'s thought where a concept is defined by its most central, or \textit{prototypical}, instance and the association of a separate instance to that concept depends on its distance to the prototype. We illustrate these ideas with the following example. A common concept is that flowers are beautiful and fragrant. However, the reality is more complex: while most flowers fit the common concept, some flowers are odorless, and some are even pungent and cause allergies. Therefore, beauty and fragrance, while brought together in the common concept of a flower, can also be taken apart. That is the case for the concept of flowers that are beautiful but pungent (or odorless). Such flowers remain semantically close to the prototypical instance (a rose, for example) of beautiful and fragrant flowers because of their many shared features, but they must be represented by a different sub-concept since, while still beautiful,  they are not fragrant. Similarly, flowers that cause allergies must be represented by yet a new sub-concept that intersects with the concepts of beautiful and fragrant flowers and beautiful but pungent (or odorless) ones. %From this, by the way, a new concept can be derived, such as the concept that not all flowers are wonderful. 
Our work equates clusters of posts to concepts and subconcepts and assesses the relations between them by means of their distance in LLM embedded space.

\subsubsection*{Lifelong Unsupervised Learning}
The method designed in this paper is inspired by a recent approach for lifelong anomaly detection presented in \cite{Corizzo2022CPDGACP, Faber2023VLADTV}. The main purpose of that approach was to build an evolving representation and organization of memory to ensure that new concepts could be coherently added to the organized structure while older pieces of knowledge could be retrieved efficiently in a lifelong learning regimen. The stability of the method ensured that catastrophic forgetting was avoided. Anomaly detection was performed on a continual basis by checking whether the new data fell into one of the categories present in memory. If not, a decision was made as to whether the new data was anomalous or not. If it was deemed anomalous, an alert was issued and the data not saved. Otherwise, the new data was introduced into the memory structure as either a sub-concept or a new concept altogether. Decisions were made based on distance in an autoencoder-based embedding space. The approach was tested and validated on tabular data representing fairly straightforward hierarchically organized categories of data. One of the challenges of the present study is to see to what extent the principles used in the earlier work can apply to text data and imprecise concepts that are not always hierarchically organized. 

\subsubsection*{Psychological Studies on Hate Speech}
 To underline the importance of the task we address with our method, we review two related psychological studies on hate speech that seek to extract similar types of information as ours from their analysis, albeit manually. In the first study, closest to our proposed one, \cite{Rohlfing2016WhoIR} extracted four semantic themes from a series of comments responding to a Youtube video that displayed a white British woman abusing black passengers on the Tube (subway). The four themes they organized the posts around are: 1) posts that explain the abuser's behavior; 2) posts that counter the abuser's behavior with aggressive and hateful language; 3) posts that categorize other posters into racist or non racist people; 4) posts that discuss what it means to be British. The authors' analysis led them to conclude that online hate speech does not necessarily generate more hate speech, a potentially useful conclusion for this specific content and platform. In a second %similar but behavioral rather than semantic 
 study, \cite{Megersa2023SocialMU} analyzed the behavioral roles taken by online users responding to hate speech posts in Ethiopia. They identified five major roles---trolling, pace-making, peace-making, informing, and guarding---and used their conclusions to make recommendations on how to counter online hatred. 
 Though like in our work, both studies support the need for a deeper semantic analysis of posts rather than their mere classification as hateful or not, it is important to note that both were conducted through human analysis of a relatively small and static number of posts. Our work, in contrast, seeks to perform the same type of analysis %done in the first study fully 
 automatically and dynamically. % using unsupervised learning and advanced NLP tools with the aim of processing large amounts of posts efficiently. 

\subsubsection*{Information Extraction from Customer Surveys}
Although the aim of our study is new in the context of automated hate speech analysis, the business community has understood the value of extracting information from product reviews for some time. As a matter of fact, it has already seized upon the power of AI to learn how to extract automatically the areas of satisfaction and concern about its products expressed by customers. This is shown in both academic research papers \cite{Ramaswamy2018CustomerPA, Aldunate2022UnderstandingCS} and commercial products available to businesses. It is important to note, however, that the domain explored by such business tools is much simpler than the hate speech domain. Indeed, human beings typically have many conflicting and hard-to-express resentments about other human beings, and these feelings may fluctuate as a result of national or international news items or personal experiences. On the other hand, feelings about purchased goods are rather straightforward and limited since they are non-conflictual and pertain to simple and finite properties such as the product's price, functionality, or looks. In addition, the approaches considered in these works take a static rather than a dynamic view of the problem.
For these reasons, the techniques used by the business community, though relevant, are not sufficient, and new methods must be developed to handle the monitoring of complex themes that also evolve over time.

\section{Background on Computer-Based Hate Speech Analysis}\label{tab:literature}
 
%The internet provides a platform for users to express themselves in ways they might avoid in face-to-face interactions. Therefore, online comments (tweets, messages) tend to be more likely to contain hateful content than real-life conversations~\cite{suler2004online}.
%Most of the work conducted recently is in the area of classification. A few surveys of hate speech detection were written in the last six years including \cite{Schmidt2017ASO, Fortuna2018ASO, Poletto2020ResourcesAB, Jahan2021ASR}. From these surveys, it is clear that there are many issues surrounding the area of hate speech detection. These include the bias in handling hate speech due to its subjective nature \cite{Garg2022HandlingBI}, the fast, ever-changing face of hate speech \cite{Qian2021LifelongLO, Israeli2022GoingEC}, and its potentially subtle manifestation \cite{Elsherief2021LatentHA}.

%We are not the only group of researchers using Deep Learning (DL). 
As previously mentioned, hate-speech detection has become an important area of study in NLP as reported by a number of surveys on the topic %large number of studies have been devoted to automatic hate speech detection 
\cite{Schmidt2017ASO, Fortuna2018ASO, Poletto2020ResourcesAB, Jahan2021ASR}.
Recently, many of these studies have employed deep learning techniques~\cite{vaswani2017attention}. In \cite{founta2019unified}, the authors proposed a Recurrent Neural Networks (RNNs) based framework
for the classification of racism \& sexism, offensive speech, and cyberbullying using text and metadata. In \cite{serra2017class}, the LSTM (Long Short-Term
Memory) architecture is used for the classification of abusive content on social media  while ~\cite{parikh2021categorizing} uses more advanced recurrent neural network techniques to classify sexism and misogyny. %Similarly, in the same context, 
In another study, \cite{badjatiya2017deep} compared multiple deep learning techniques to classify tweets into three categories: racism, sexism, or neither. 

There have been a few studies devoted to the particular issue of online antisemitism. \cite{chandra2021subverting} used a transformer architecture (RoBERTA) to classify antisemitic content into four categories: political, economic, religious, or racial. In \cite{barna2021exploration} the authors used Latent Dirichlet Allocation (LDA) to find topics particularly related to the coronavirus and antisemitism in the Hungarian language. %. They analyzed articles and comments written in Hungarian. While that work is related to ours, LDA requires that the number of expected topics be known before running the model. The authors worked around this requirement by using a topic-coherence metric \cite{roder2015exploring} to decide on the number of topics appropriate for the given corpus. Here, we argue that this approach is only feasible for particular themes such as coronavirus and antisemitism. In our work, we stay clear of methods that require advanced knowledge of the number of topics to detect.
\cite{zannettou2020quantitative} describes a large study with several outcomes. In that work,  the authors collected 7 million images and comments from
\texttt{4chan} and \texttt{Gab} to study the spread of antisemitic
memes. They provide a framework to better understand antisemitic memes and content and their relation to political events. In one part of the study, they used Word2Vec to discover words closely associated with the term "Jew" and used network analysis tools such as community discovery in an attempt to uncover the themes represented by these terms. The themes they uncovered are: Jewish moral corruptness, Jews as geopolitical conspirators, ethnic Jewish identities, Kabbalistic and cryptic themes, and religious or theological topics. Following this study, \cite{ali2022analyzing}
%The work closest to ours in purpose but not methodology 
considers both antisemitism and Islamophobia. %\cite{ali2022analyzing}. 
In that work, the authors constructed two lexicons: a lexicon of 48 Antisemitic terms and another of 135 Islamophobic terms. The lexicons were built using pre-existing knowledge graphs, pre-trained word embedding models, and manual annotations. Using their lexicons, they extracted posts from 4Chan and analyzed them using sentence embeddings by BERT and clustering using HDBSCAN. They subsequently analyzed the groupings qualitatively. %and assessed their popularity and veracity (i.e., what percentage of posts that contain these terms are actually Antisemitic/Islamophobic). 

%Antisemitism has been widely studied in the social science domain (e.g., \cite{salwen2009antisemitic,schwarz2017inside}). These studies provide the history behind antisemitic content and hate against Jews. Hate against Jews is not a new phenomenon, but the growing trend on social media has created many difficulties in recent times.  In \cite{zannettou2020quantitative}, the authors collected 7 million images and comments from
%\texttt{4chan} and \texttt{Gab} to study the spread of antisemitic memes. They provide a framework to better understand antisemitic memes and content and their relation to political events. In contrast to this work, we worked on two known antisemitic issues using state-of-the-art deep learning techniques, i.e., i) antisemitic tropes discovery and ii) antisemitic coded terms floated on extremist social media. We provide solutions to both tropes and terms using deep learning techniques.
%Moreover, traditional approaches to hate detection do not provide solutions to more subtle problems such as detecting and monitoring the specific antisemitic themes that appear online (e.g., Jews are greedy, Jews killed Jesus, Jews are disloyal, etc. ) and identifying the coded language often used to express these themes (i.e., new world order, globalist, moon landing, etc.). 

We summarize the previous work in %the domain of hate speech detection in 
Table~\ref{tab:comparison_table}. Our proposed approach is summarized at the bottom of the table. 
As can be seen, our work is most similar to \cite{barna2021exploration}, \cite{zannettou2020quantitative}  and \cite{ali2022analyzing}. The main difference between these three works and ours is that they consider a static situation whereas our approach is an online machine learning approach that provides self-adaptation mechanisms in a dynamic setting, making sure that the knowledge base induced in previous steps is not compromised by a new influx of information. Self-adaptation, dynamism, and stability aside, of the three methods closest to ours, \cite{ali2022analyzing} is the one that uses the closest methodology since \cite{barna2021exploration} performs topic extraction using LDA, and \cite{zannettou2020quantitative} does so using Word2Vec. These two methodologies are less recent and typically obtain lower performance than the one used by both \cite{ali2022analyzing} and our work: transformers. In particular, both works use BERT, though while \cite{ali2022analyzing} rely on pre-trained language models, we fine-tuned our model to the particular type of data we analyze. Like in our study, \cite{ali2022analyzing} uses clustering on the language representations learned and, in particular, HDBSCAN. Rather than using an off-the-shelf static clustering approach, however, we design our own approach that works in a dynamic environment without causing catastrophic forgetting.
\begin{table*}
    \centering
        \caption{Comparison to traditional work and techniques used in hate speech detection.}
    \begin{adjustbox}{width=\textwidth}
    \begin{tabular}{|ccccccm{5cm}|}
    \hline
         Ref& Unsupervised & Online &Term Extraction & Topic Extraction & ML Tools & Hate Type\\\hline
         \cite{founta2019unified} & $\times$ & $\times$ &$\times$ &  $\times$ &  RNNs& racism \& sexism, offensive speech, and cyberbullying\\\hline
        \cite{serra2017class} & $\times$ & $\times$ &$\times$ & $\times$ & LSTMs & abusive content \\\hline
        \cite{parikh2021categorizing} & $\times$ & $\times$ &$\times$ & $\times$ & LSTMs, CNNs & sexism and misogyny \\\hline     
        \cite{badjatiya2017deep} & $\times$ & $\times$ &$\times$ & $\times$ & LSTMs, CNNs, GloVe & racism, sexism, or neither \\\hline
        \cite{chandra2021subverting} & $\times$ & $\times$ &$\times$ & $\times$ & RoBERTa & antisemitism %content and category (Political, Economic, etc.) using text and images 
        \\\hline
        \cite{barna2021exploration}& \checkmark & $\times$ &$\times$ & \checkmark & LDA & coronavirus related antisemitism %in Hungarian language  
        \\\hline
        \cite{zannettou2020quantitative}& \checkmark &$\times$ & \checkmark & \checkmark & Word2Vec, network analysis& antisemitism\\\hline
        \cite{ali2022analyzing}& \checkmark & $\times$ &\checkmark & \checkmark & BERT, HDBSCAN & antisemitism and islamophobia\\\hline
        This work & \checkmark & \checkmark & \checkmark &\checkmark & BERT, novel algorithm & antisemitism %, applicable to other kinds of hatred, such as Islamophobia, anti-black or Asian racism,
%anti-LGBTQ+  
\\\hline
    \end{tabular}
    \end{adjustbox}

    \label{tab:comparison_table}
\end{table*}

\begin{figure}[tp]
\vspace{-2mm}
   \centering
   \includegraphics[scale=0.5]{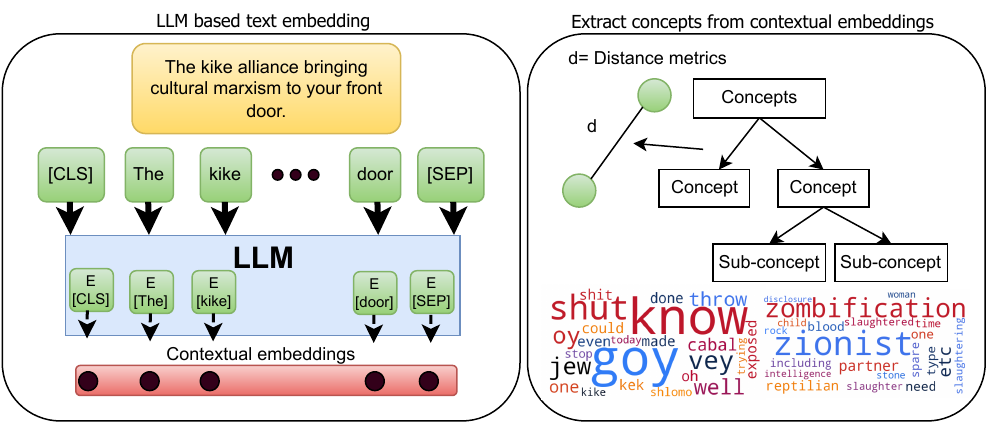}
   \vspace{-3mm}
   \caption{Overview of the methodology.}
   \label{tab:ytfig}
   
\end{figure}
%\section{Methodology} \label{tab:methodology}

\section{Methodology} \label{tab:approach}
Our antisemitic discourse monitoring approach takes as input continuously arriving batches of posts and organizes them into a knowledge base containing the antisemitic themes they convey in an unsupervised way. In more detail, we designed a repeating two step-method that takes the following actions for each batch of incoming posts: 1) it generates contextual embeddings from the posts using a fine-tuned BERT model; 2) it inputs these embeddings into our newly proposed divisive clustering approach which adaptively updates the last clustering to integrate the new batch of embeddings into the existing knowledge base using local or global updates. %Existing clusters representing antisemitic themes are either updated to include new posts similar enough to those they represent, or they are divided into sub-clusters that collectively closely model  all the antisemitic themes encountered in the data so far. The decision to update a cluster or divide it is 
Decisions are made based on the distance in embedding space between the new posts and the existing cluster centers. The two main mechanisms used in the process are illustrated in   Figure \ref{tab:ytfig}. Each step of the process will be described in more detail below.

%In this section, we describe the way in which LLMs are integrated into our antisemitic themes discovery approach. LLMs typically tokenize text into smaller units, like words or subwords, ensuring that the input data is tokenized in a compatible way. The flow of the approach is depicted in Figure \ref{tab:ytfig}. The initial step of our approach, not shown in the figure, involves preparing the dataset. Afterward, we feed it into LLMs for tokenization, generating contextual embeddings that facilitate the discovery of underlying concepts. Once concepts are discovered, we extract terminology associated with these concepts to help users of the system quickly identify and classify the kind of posts they encounter. This section is divided into four subsections. We first describe the data set and its pre-processing, followed by the context embedding method we used. Next, the algorithm for trope extractions is described and we conclude with an explanation of our term extraction process.

\subsection{Dataset and Pre-Processing}
%Therefore, we follow the pre-processing steps below during the first step of converting the raw text into information.

\subsubsection{Data Collection and Labeling} This work is part of a larger project entitled  \texttt{unmasking antisemitism} that our organization has undertaken to control anti-Jewish hatred. Under this project, a data team collected and labeled the data set using third-party software Pyrra\footnote{https://www.pyrratech.com/}. A baseline was set for data selection under which currently known %, both coded and non-coded, 
antisemitic seed words from the American Jewish Committee (AJC) and other sources were used (e.g., cabal, new world order). During labeling, the data team carefully examined word selection in posts as well as other pieces of information such as i) does the post demonstrate an applied understanding of particular antisemitic tropes? ii) does the post express themes of control over media, politics, academia, economy, or culture at large? and so on. It then classified the data into different pre-defined historic tropes or generated variations on them when the post was found to be antisemitic. % The data team also considered a few other checks while labeling, for example, i) Does the post demonstrate an applied understanding of antisemitic tropes?, ii) Does the post express themes of control over media, politics, academia, economy, or culture at large? and so on. 
The labels generated by the data team are used in this work to evaluate our approach and the baselines we will be comparing it to.%A few more criteria were set, which are in the domain of the data team.
\footnote{The coding statement used by the data team is available upon request.} %However, a brief explanation is available on this link \textbf{LINK or Appendix}.
\subsubsection{Data Pre-Processing}
The initial step of our approach, not shown in Figure~\ref{tab:ytfig}, involves preparing the dataset using different techniques. We first normalize the text documents by converting uppercase to lowercase and replacing references. Next, we remove links from the text. However, we replace the links with the titles of the posts they lead to. 
% We also remove {\color{red} ambiguities PLEASE SEE COMMENT AND EXPLAIN} in the text. 
Finally, we remove all special characters, white spaces, and line breaks using regular expressions. %In addition, LLMs typically tokenize text into smaller units, like words or word parts. 
The cleaned-up texts are then fed into BERT for tokenization into small units (words or word parts) and contextual embeddings are generated to facilitate the discovery of underlying concepts in the text. The details of the contextual embedding method used here are described in the next section. 
 \begin{figure}[tp]
\vspace{-2mm}
   \centering
   \includegraphics[scale=0.6]{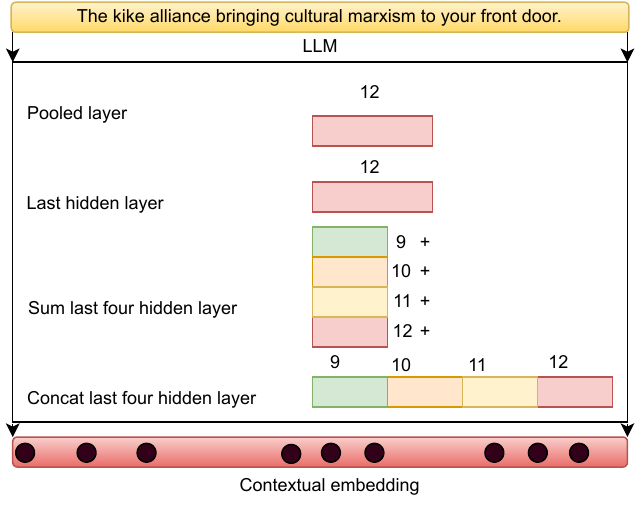}
   \vspace{-3mm}
   \caption{BERT layers to extract contextual embedding.}
   \label{tab:afig}
\end{figure}
\subsection{Contextual Embedding} \label{tab:embbeding}
To convert our extremist cleaned-up dataset into embeddings, we use embedding techniques derived from large language models to extract a structural representation that captures semantic relationships and contextual meanings. This approach facilitates the categorization of information to uncover hidden patterns \cite{chandra2021subverting}. Since BERT doesn't possess vocabulary specific to our domain, we first added Out of Vocabulary (OOV) tokens and fine-tuned the BERT uncased model as in ~\cite{Kikkisetti2024UsingLT}. %Such tokens are broken down into smaller units or subwords yielding unsatisfactory results.
During concept extraction, we consider different types of combinations to extract contextual embedding from the BERT (see Figure \ref{tab:afig}). To this end, we first explored the summation of the last four layers, followed by the concatenation of the last four layers. We also considered the pooled layer, which provides 768-dimensional sentence-level embeddings. In our case, the last hidden layer, after taking the mean of tokens level embedding, produces the best results and was therefore the embedding technique we retained. These variations allowed us to evaluate the impact of different feature representations, contributing to a deeper understanding of the model's capabilities and suitability for our specific NLP task. 

\subsection{Antisemitic Themes discovery} \label{tab:Tropes_approach}
We now describe the approach we designed to extract concepts from extremist posts. To simulate a streaming situation, we divide the dataset into batches of vectors $\vec{v}$ (each containing instances of posts embeddings) and feed each batch into algorithm \ref{alg:two}. Algorithm \ref{alg:two} uses distances in the embedding space along with several thresholds to capture the antisemitic themes expressed in our extremist dataset. Several parameters need to be set prior to running the algorithm: $k, lo, hi,$ and thresholds $ \lambda$ and $\delta$. $k$ represents the number of initial concepts considered in the first batch of data $\vec{v_1}$. In our work, we set $k = 2$, assuming that the posts in $\vec{v_1}$ align along two main themes. $lo$ and $hi$ represent the lower and higher percentile marks used in our computations to consider whether a post is within the vicinity of a concept or not. We set them to $lo = 40$ and $hi = 60$, here. $\lambda$ is the multiplier used (combined with $lo$ and $hi$) to calculate the percentile distribution ranges that will help determine whether instances from a given batch are outliers with respect to the concept considered. Finally, $\delta$, determines the minimum number of outlying instances required to justify the need for an additional concept in memory (i.e., an increase in $k$). {In our work, we choose a $\lambda$ value of 0.25, and $\delta$ is set to 15\% of the batch size.} 
% {\color{red} Please indicate what values of lambda and delta you used to run the algorithm.}

\begin{figure}
    \centering
    \includegraphics[width=1.1\linewidth]{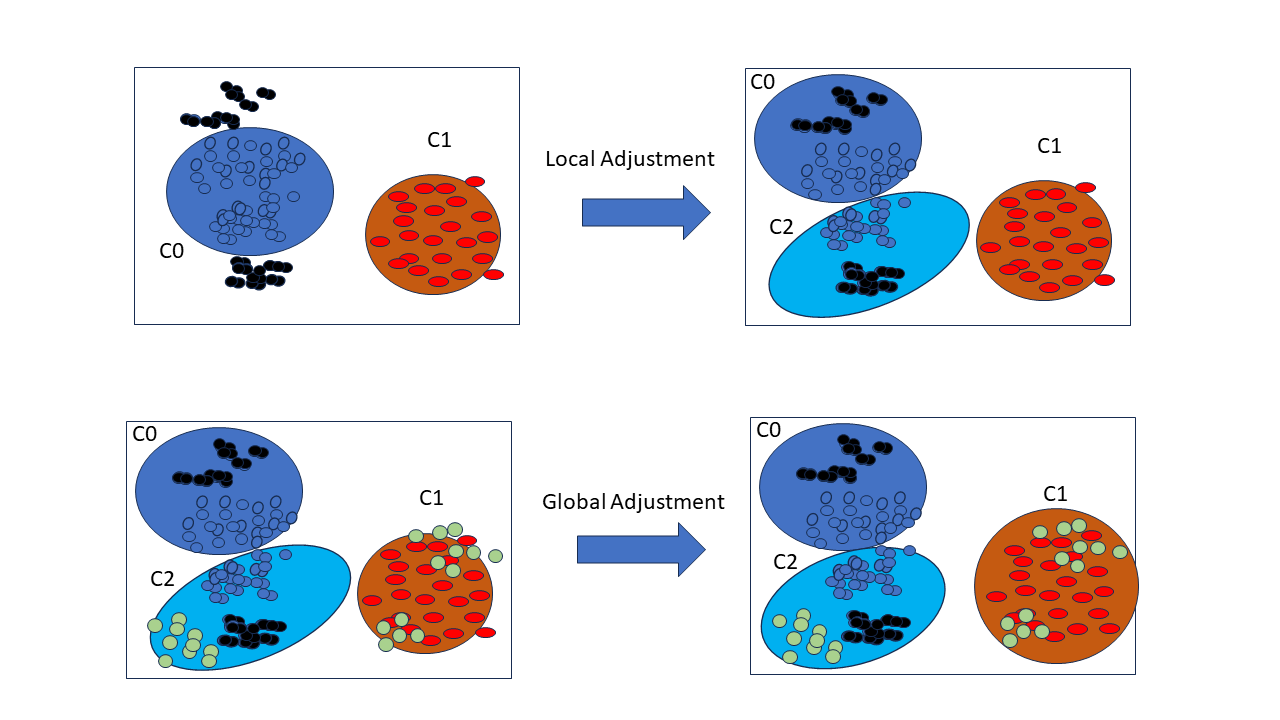}
    \caption{Local and Global Updates. }
    \label{fig:algo-explanation}
\end{figure}

%deviating from a concept already in the memory which  will allow the algorithm to determine We consider $k$ concepts to start this continual process of extracting concepts. Next, we have threshold {$\delta$}, which is the total number of instances deviating from a concept already in the memory, and multiplier $\lambda$ used in percentile distribution ranges. 
A crucial aspect of the algorithm is its stability. Indeed, given its continual learning nature, it is imperative that when new information comes in, previous knowledge is preserved to the greatest extent possible. % while the new information gets integrated. 
To implement this requirement, we only allow for two kinds of updates: a local update and a global one, each with the following allowances and limitations.  
%
%a new clustering is not learned from scratch. Instead, 
%with local updates possible through the eventual splitting of single concepts into pairs of related ones, and global updates to the whole structure. This means that we will need a local adjustment phase and a global adjustment phase. 
The local adjustment phase may be drastic, but it can involve only one concept. The global adjustment phase is much lighter %in that only small changes to current clusters are allowed, 
but it may affect all the concepts in memory. The two adjustments are shown in Figure~\ref{fig:algo-explanation} where the top figure illustrates a local adjustment while the bottom one illustrates a global one. The leftmost pictures show the ``before" situation with black dots on the top figure and green ones on the bottom representing data that has arrived in a new batch, but has not yet been processed. The rightmost pictures show the ``after" situation, i.e., the result of the processing in each case. What is important to note in the top figure is that while the original C0 concept gets split into two new concepts C0 and C2, concept C1 is not affected by this change. That is an instance of a local adjustment. In the bottom figure, an example of a global adjustment, no new concept is created, but both C1 and C2 are slightly modified (their center and/or spread change from the picture on the left to that on the right) to accommodate the new data. C0 could have accommodated new data as well, but in our particular illustration, no new data affected it. 
%and includes cluster centers to move slightly from their current location.

We now describe Algorithm~\ref{alg:two} line by line. It begins on line 1 where the batch number $b$ is set to 1. On line 2, a model is built by clustering the data in $\vec{v_1}$ using the k-means algorithm (we chose an initial value of $k = 2$).  Next, on lines 3-4, we store the initial centroid values in $cc$ and initialize the semantic similarity structure, $ss$, that associates the two concepts most closely related to one another. Specifically, $ss$ is a list of $[r, l]$'s, where $r$ refers to the root concept that already existed in memory, and $l$, the concept derived from $r$ during local updating. $l$ has some degree of similarity with $r$ though the majority of the data in $l$ represents a new concept. On line 4, we have $r_1 = r_2 = 0$ since these are the initial concepts with no root from which tehy derived.  
%
%and also keep track of hierarchies during the whole process. On line 2, We train the initial batch of extremist text vectors $\vec{v}$ on line 4. Next, on lines 5-6, we store the centroid values and also keep track of hierarchies during the whole process. We initialize a list with tuples to keep track of new concepts and their root concept. 
During our continual learning process, new concepts are created or old ones modified based on percentile distributions and numbers of outlying instances, as illustrated in Figure~\ref{fig:algo-explanation}. %deviating from $\delta$. 
On line 5, we increment the batch number and on line 6, we assign the next batch of posts to $\vec{curr}$. Line 7 is the while loop that drives the continual consideration of new batches of posts. Inside the loop, line 8 calculates the size of the batch and assigns it to $s$. Line 9 is another loop that considers each of the concepts currently present in memory, 
 in turn. The purpose of that loop is to implement the local updates of the type shown on the top of Figure~\ref{fig:algo-explanation}. 
In more detail, lines 10-22 are repeated for each concept in memory with respect to the current batch, $\vec{curr}$. The purpose of the loop is to estimate whether the data in the batch fits the current concept well or whether it would be beneficial to split it. This question is repeated for each of the current concepts in memory and in every case, if the answer is yes, $k$ gets incremented. %This means that given a new batch, if $t$ concepts are currently present in memory, $k$ may be incremented by anywhere between 0 and $t$. 
The decision of whether to increment $k$ or not for a given concept is done by first, determining the lower and upper percentiles of the distance distribution to find out whether a particular post belongs to concept c and falls within the outlier category, and second, seeing whether the number of instances from the batch to be considered outliers by these percentiles exceeds a specific threshold. In more detail, on line 10, a vector is calculated that stores the distance of each instance in the batch to the concept's centroid. Lines 11-12 calculate the lower and upper distance values representing the $lo^{th}$ (40) and $hi^{th}$ (60)'s percentiles. These correspond to the limits within which a post is considered the purview of concept c. These values are used on lines 13-14, along with threshold $\lambda$ (.25) to determine the lower and upper percentile distribution ranges used to establish whether an instance in the batch is an outlier or not.   On line 15, $\vec{out}$ is assigned all the instances from the current batch that are the purview of concept c and yet deviate too much from its centroid (either from the lower or from the upper end). On line 16, the number of instances in $\vec{out}$ is pitted against threshold $\delta$. 
If that number exceeds $\delta$ (15\%), k-means is run with k=2 on all instances of concept c ($\vec{c}$) and all instances contained in $\vec{out}$ on line 17 to split the concept and its outliers appropriately (see top of Figure~\ref{fig:algo-explanation}). On lines 18-20, the centroids cc are updated, the number of concepts present in memory, $k$, is incremented, and the cluster pair is added to the semantic similarity list, $ss$. %, increment $k$ (labels) to model and update cluster centroids for all concepts. This is done to ensure that the entire set of clusters reflects the most recent information from the entire dataset.
% 
% Once the process of estimating $k$'s optimal value for all the concepts currently in memory has ended, k-means is run (with the new $k$) on all the data seen so far including the data in $\vec{curr}$. The centroids and hierarchical structure are then updated on lines 19-20 according to the new clustering. 
%The way this pair is calculated for each centroid c is by searching for the cluster closest to c. 
For example, in the top situation of Figure \ref{fig:algo-explanation}, C2 is created and was derived from C0 (which itself was shifted), so we add $[C0, C2]$ to $ss$. 
Line 21 ends the ``if statement" that causes a local update to occur, while line 22 terminates the inner loop that considers each concept in memory one after the other. Lines 23-24 implement the second type of adjustment, the global adjustment, that takes place in our algorithm. This consists of simply running k-means on the union of all batches of data seen so far with the updated value of $k$. Since $k$ fits the data perfectly thanks to our carefully crafted local adjustment approach and since k-means has theoretical guarantees of stability \cite{BenDavid2007StabilityOK}, line 23 will have the minimum effect of adjusting the centroids as shown at the bottom of Figure~\ref{fig:algo-explanation}, thus adjusting the knowledge in a stable manner, as required. Lines 25-26 increment the batch number and assign the next batch to $\vec{curr}$ prior to looping back to line 7. The algorithm terminates if no new batch is present.
It is important to note that this process is a continual learning process. However, though our approach has focused on the stability of the memory-building process, it has left out a number of features necessary in lifelong learning. %; if we come up with a new extremist dataset, the algorithm will consider it using already available concepts in memory and, if required, create new concepts. 
To ensure our system's long-term viability, we will outfit it, in the future, with lifelong learning features such as those proposed in \cite{Faber2023VLADTV} including knowledge distillation and experience replay.
\begin{algorithm}[hbt!]  
\caption{Antisemitic concepts extraction\\
Input: $\vec{v} = \begin{bmatrix}\vec{v_1}, \vec{v_2},...,\vec{v_n} \end{bmatrix}$ ; %\Comment*[r]{Batch embeddings}\\
Parameters: $k$, $\delta$ , $\lambda$, $lo$, $hi$}
\label{alg:two}
%\Comment*[r]{$\delta$\ sets t}
% \Comment*[r]{sensitivity $\lambda$}}\label{alg:two}
% \KwData{$n \geq 0$}
% \KwResult{$y = x^n$}
% $b \gets x1$\ \Comment*[r]{batch size}

%$\vec{v} \gets \begin{bmatrix}\vec{v_1}, \vec{v_2},...,\vec{v_n} \end{bmatrix} $\Comment*[r]{Batch embeddings}
% $c \gets 2$\ \Comment*[r]{initial concepts}

% $K \gets c $ \Comment*[r]{initialize model with c}
$b \gets 1$ \Comment*[r]{initial batch number}
$m \gets kmeans(\vec{v_b},k)$ \Comment*[r]{build initial model}
%$c \gets concepts $ \Comment*[r]{concepts in the memory}
$cc \gets m (k) $ \Comment*[r]{$cc$ contains centroids $c_1,..., c_k$}
$ss \gets [(r_1, l_1),..., (r_k, l_k)]$ \Comment*[r]{initialize semantic similarity structure} %with root/centroid information}% for initial concepts. $r_1 = ... = r_k = 0$,% since the two initial concepts have no parent
$b \gets b + 1$\Comment*[r]{increment batch number}
$\vec{curr} \gets \vec{v_{b}}$ \Comment*[r]{Consider next batch}
\While{$\vec{curr}$ != Empty}{
$s \gets len(\vec{curr})$ \Comment*[r]{numb. of instances in $\vec{curr}$}
      \ForEach {$concept$ $c=1$ $\textit{to}$  $k$ \textit{in} $memory, m$}{
    $ \hat{d^c} \gets \begin{bmatrix}\hat{d_1^c}, \hat{d_2^c},...,\hat{d_s^c} \end{bmatrix}$ \Comment*[r]{pred. dist. to c}
    $l \gets percentile(\hat{d^c},lo) $ \Comment*[r]{$lo^{th}$ perc. in $\hat{d^c}$}
    $u \gets percentile(\hat{d^c}, hi)$ \Comment*[r]{$hi^{th}$ perc. in $\hat{d^c}$}
    $lower \gets l - \lambda * (u-l)$ \Comment*[r]{lower percentile}% distribution}
    $upper \gets u + \lambda * (u-l)$ \Comment*[r]{upper percentile}% distribution}

  $ \vec{out} \gets (l< \vec{curr} < \textit{lower}) + ( u > \vec{curr} > \textit{upper})$ \Comment*[r]{ set of outliers}
    \If{$len(\vec{out}) > \delta$}{
        $kmeans([\vec{c}, \vec{out}], 2)$ \Comment*[r]{local adjustment}
        % $c+=1$  \Comment*[r]{increment numb. of concept}
        % $update$ $labels$ $l$ \Comment*[r]{upd. labels}
        $update$ $cc$ \Comment*[r]{add new centroid to cc} 
        $k+=1$  \Comment*[r]{increment numb. of k}
        $ss \gets (r_c,l_c)$ \Comment*[r]{root and child concept}
    }
    }
    $m \gets kmeans([\vec{v_1},...,\vec{v_b}],k%-\vec{out}
    )$ \Comment*[r]{global adjustment}
    $cc \gets m (k) $   \Comment*[r]{update all centroids}
       % $h \gets [(r_1, c_1),..., (r_k, c_k)]$ \Comment*[r]{update h}    
        $b \gets b + 1$\Comment*[r]{increment batch number}
$\vec{curr} \gets \vec{v_{b}}$ \Comment*[r]{Consider next batch}
%$\vec{curr} \gets next(\vec{v})$ \Comment*[r]{Consider next batch}
  }

  \end{algorithm}

\vspace{-.25cm}
\section{Results and Discussion} \label{tab:results_and_discussion}
%The system we propose is difficult to evaluate given its novelty, which means that there are no simple baselines to compare it to; the double tasks it performs--trope discovery and terminology extraction; and its exploratory nature, which means that the output of the system cannot be pit against any ready-to-use, nicely-compiled body of knowledge. 
%To address these challenges, we conducted several types of evaluations, some quantitative and others qualitative. 
%The system we proposed performs two major tasks---trope discovery and terminology extraction. We discuss their respective evaluation in turn. We conducted several types of evaluations, some quantitative and others qualitative. %how we evaluated the trope discovery aspect of our approach. We then discuss how we evaluated the term extraction aspect. 
%In the term extraction part, in particular, we propose a new automated method of evaluation that we pit against a more cumbersome but human-controlled, manual one. We conclude the section with a discussion of two additional issues: the dynamic building of a knowledge graph and the handling of neutral terms.

%\subsection{Antisemitic Tropes Discovery}
We now discuss both the quantitative and qualitative ways in which we tested our trope discovery approach. %It then discusses the proximity analysis used to output our graph of trending tropes.
We recall that our approach is dynamic and creates clusterings over time. Unfortunately, there are no clear baselines for comparing the performance of our approach over others in that fashion. Therefore, we consider the final clustering output by our approach and compare it quantitatively to those of state-of-the-art static clustering methods, along with qualitative considerations. %The dynamic aspect of our approach will be considered qualitatively in Section~\ref{sec:addcons}.  
The section starts with a quantitative analysis, followed by a qualitative one. Using specific information from the qualitative analysis, we then return to the quantitative analysis focusing on particular regions of the clustering.
\subsection{Quantitative Analysis}\label{sec:quant-an}
We compared the final clustering obtained by our method to the following five state-of-the-art clustering approaches.\\
%\begin{itemize}
    %\item 
    \textbf{Affinity Propagation}: Affinity Propagation groups data points together relying on a message-passing process between points that assesses their similarity to the different groupings. %based on similarity. It also ensures that each data point has the highest similarity within a group. 
    The algorithm doesn't require pre-defining the number of clusters~\cite{dueck2007non}.\\
%    \begin{itemize}
       % \item Affinity propagation doesn't require pre-defining the number of clusters compared to K-Means~\cite{dueck2007non}. This approach groups together data points based on similarity. It also ensures that each data point has the highest similarity within a group. Using this approach, we obtained 31 clusters that did not represent tropes categorization in any sense.  Therefore, in our extremist dataset, this approach fails when we experience multiple times a single sentence makes a complete concept. Nevertheless, we have a much higher number of clusters representing the distributions; the evaluation metrics are still not promising see Table \ref{tab:baseline_results}.  
      %  In our approach, concepts with contextual similarity are grouped together to make it a single concept, i.e., \texttt{Shut it down the goyim know}. This concept presents 15.90 \% of the dataset. Using the Algorithm~\ref{alg:two}, we group together with 75\% of the concept \texttt{Shut it down the goyim know} and 100\% the extremist concept (Slaughter of Jews) in a single concept. However, using Affinity propagation, C2 splits into two tropes providing  30.50\% and 23.72\% accuracy, while this approach correctly classifies C7.
 %   \end{itemize}
    %\item 
    \textbf{Birch}: Birch begins by producing a brief summary tree structure that preserves the maximum amount of information in the dataset. In the next step, this summary is classified into different numbers of clusters. Birch can be used with or without specifying the desired number of clusters, $k$. We ran it without specifying, $k$ and obtained over 300 clusters. We then searched for the $k$ that optimized the results and settled on $k=9$ \\ %  Using Birch, we select different combinations of clusters to compare the results with our approach.
       % \begin{itemize}
         %   \item Birch first produces a brief summary that preserves the maximum information of the original dataset. In the next step, this summary is classified into different numbers of clusters.  Using Birch, we select different combinations of clusters to compare the results with our approach. First, we set the number of clusters, \texttt{n\_cluster=None} and the algorithm provides >300 tropes. None of the tropes make any sense, and the use cases (C2, C7) we defined for qualitative analysis are overlapping to each other. Next, we try different combinations of $k$ without any clear separation of the tropes. Our method yielded a maximum of 9 tropes, which clearly present extremist text with similar semantic meanings. When using $k=9$, Birch correctly classifies 67.79\% of trope C2 with nearly the same accuracy for C7. Moreover, we also maintain a record of semantically proximate concepts $[r,c]$ along with the facility to extract concepts in a continuous fashion from the extremist dataset.       
       % \end{itemize}
%\item 
\textbf{Spectral Clustering}: Spectral Clustering is derived from spectral graph theory. It involves analyzing the eigenvalues and eigenvectors of a similarity matrix derived from the data. The number of desired clusters, $k$, must be fixed prior to running the algorithm. We considered different combinations of parameters, %defined in the scikit-learn package, 
but none of the combinations clearly differentiated between the antisemitic themes. For comparison purposes, %However, to compare the results with our approach, we 
we settled on $k = 9$.\\ %consider 9 tropes for this approach.\\
%\begin{itemize}
%    \item  Spectral Clustering derived from spectral graph theory. It involves analyzing the eigenvalues and eigenvectors of a similarity matrix derived from the data. Using this approach, we consider different combinations of parameters defined in the scikit-learn package, but none of the combinations clearly differentiate the tropes. However, to compare the results with our approach, we consider 9 tropes for this approach. We obtained one concept that considers more than 81\% of data in a single cluster. However, the rest of the $8$ concepts consider only 5\%. 
 %   For example, using our approach, we successfully made a single concept of the meme \texttt{Shut it down the goyim know} -- C2 with an accuracy of 75 \%. However, using spectral clustering, the extremist posts were distributed into more than one concept. %see Table \ref{tab:otherCls}. 
  %  Similarly, C7 is merged with C4, and there is no clear separation between them. 
   % \end{itemize}
    % In our use-case, we set $n_clusters=9,n_neighbors=10$.
    %\item 
    \textbf{Gaussian Mixture}:
    %\begin{itemize}
        In Gaussian Mixture (GM), each cluster is represented as a Gaussian distribution, and data points are assigned to clusters based on their likelihood of belonging to each %Gaussian 
        component.
        GM requires a pre-defined number of clusters, $k$. Here again, after evaluating different combinations to look at the distributions of clusters, we use $k = 9$  to compare the results to our method.\\ %$k = 9$ provided a good number of clusters for comparison purposes. \\%Therefore, in our use case, after evaluating different combinations to look at the distributions of clusters, we use 9 tropes to compare the results. Using GM, we achieve 33.89 \% of grouping together \texttt{Shut it down the goyim know} -- C2 in a single concept, while C7 has nearly the same accuracy. Furthermore, we posit similar observations to those made for the alternative approach, \textbf{Birch}.    
%\end{itemize}
%\item 
\textbf{Mean Shift}: Mean Shift is a centroid-based algorithm that updates the centroid with each new point. Mean Shift works in high-dimensional spaces and uses the mean as the centroid \cite{derpanis2005mean}. %assigns the centroid to the mean of each of the discovered clusters \cite{derpanis2005mean}.  
It does not require a predefined number of clusters. 
The results of our comparison are shown in Table~\ref{tab:baseline_results}. In this discussion, we focus on columns 1-4 of the table. The last two columns will be discussed in Section~\ref{sec:detailed-results}. We use three well-known metrics devised for unsupervised learning: the \textbf{Davies-Bouldin Index (DBI)}, the \textbf{Silhouette Coefficient (SC)}, and the \textbf{Calinski-Harabasz Index (CHI)}. The first metric, DBI, computes the average similarity between each cluster and its most similar one. In contrast to the next metrics, a clear separation is indicated by a lower score. %{\color{red} According to DBI, our method is the worst!!!!! Is that correct?} 
The second metric, SC, measures the similarity of each point within a cluster to its similarity to the nearest cluster. It then averages all the results. It returns a value between -1 and 1, where 1 represents an ideal clustering. %completely different clusters and -1 means all datapoint assigned to the wrong clusters. A positive value indicates that there is degree of separation between data points.  
The third metric, CHI, also determines the quality of clustering by measuring the ratio of between- to within-cluster dispersion, however, it does so at a more abstract level than SC. Here again, the higher the value of CHI, the clearer the separation of the data points; however, there is no upper bound. Table~\ref{tab:baseline_results} also shows the number of clusters discovered by each clustering method on its own when possible (underlined), or assigned to it. Figure~\ref{tab:tsne} in the appendix gives a visual illustration of each approach's results by displaying their T-SNE projections.
\begin{table*}
    \centering
    \caption{Comparison of our approach with five baselines.}
    \begin{tabular}{|lllllll|}
    \hline
         Approach& Davies-Bouldin & Silhouette Coefficient & Calinski-Harabasz & Clusters & C2 Coverage & C7 Coverage \\\hline
        Affinity  &  0.85& 0.34 & 407.22 & \underline{31} & 30.50 \& 23.72\% & 100\%  \\\hline
        Birch  &  0.86 & 0.35 & 384.60 & 9 and \underline{300+} & 67.79\% and <5\% & 100\% and <5\% \\\hline
         Spectral & 0.64 & -0.01 & 67.76 & 9 %and 8 
         & Not Applicable & Not Applicable\\\hline
         Gaussian & 0.81 & 0.30 & 283.65 & 9 & 33.89\% & 100\% \\\hline
         MeanShift & 0.76 & 0.45 & 394.10 & \underline{2} & Not Applicable & Not Applicable\\\hline
         Our approach & 0.86 & 0.33 & 340.38 & \underline{9} & 75.50\% & 100\% \\\hline
    \end{tabular}
    
    \label{tab:baseline_results}
\end{table*}

Putting aside the dynamic element of our approach not present in any of the methods we compared it to, these results suggest that our approach is the only one able to estimate a reasonable number of clusters while also performing near the top. %pulling a clustering algorithm off the shelf for use in this application is not the best approach. First, is the issue of $k$, the number of clusters. 
Indeed, none of the methods that discover $k$ on their own did a reasonable job: Affinity discovered 31 clusters, Birch, over 300, and MeanShift, only two.  In all these cases, the results are not very practical and disqualify Affinity and MeanShift. Birch, on the other hand, remains in the game since it allows the user to provide a predefined number of clusters. Our approach settled on 9 clusters which we use going forward. %, as will be discussed below is a reasonable number for this domain. 
Once provided with the number of clusters ($k = 9$), Birch, Gaussian, and our method perform reasonably well according to two of the metrics considered (SC and CHI) with Birch leading the pack, whereas Spectral does well on DBI and poorly on the other two metrics. A look at the T-SNE plot for Spectral reveals that the algorithm learns one very large cluster and eight very small ones, therefore, not capturing the essence of the dataset. This observation along with Spectral's very low scores on SC and CHI disqualify it from further consideration. We, thus, conclude that Birch, Gaussian, and our method are competitive for this domain with Birch possibly learning slightly better clusters as seen by its slightly better results on SC and CHI (Gaussian does poorly on CHI, a little worse than Birch and our method on SC though a little better than both methods on DBI). Nonetheless, we recall that our method has two advantages over Birch (and Gaussian): 1) it is a continual and stable self-adjusting algorithm; 2) it does not require a pre-defined number of clusters. 

\subsection{Qualitative Analysis}\label{sec:qual-an}
\begin{table*}
    \centering 
    \caption{Trending Terms present in the extracted concepts}
    \begin{adjustbox}{width=\textwidth}
    \begin{tabular}{| m{10em} | m{7.3cm}| m{7.3cm}|}  \hline
        Concepts & Unigrams & Bigrams/Trigrams\\ \hline
         C1: Control (Economic)
 &   world, new, order, evil, need, state, zionist, globalists, economic, forum&new world order, world economic forum, deep state, banking cabal, state cabal	   \\ \hline
 C2: Control (Economic \& Political) &   goyim, know, shut, jew, oy, well, kike, cope, oey, kek&goyim know, know shut, many kike, low birthrates, cunning plan, know fire	   \\ \hline
  C3: Control (Cultural) &   jewish, cultural, war, marxism, frankfurt, school, political, group, white, lobby&frankfurt school, political correctness, central bank, white reproduction, holy war, correct coalition	   \\ \hline
   C4: Control (Political)   &  jewish, war, military, nation, money, control, Ukraine, russia, communist&mel gibson, zionist partners, zionist agent, ethnic foods
, interest group, jewish community, moon landing, deception mentality, communist coup, color revolution	   \\ \hline
    C5: Religion&   zionist, type, Jesus, reptilian, blood, time, know, f**k, people&blood type, zionist reptilians, jews bible, real jew, fearsome warrior, rabbinical jew	   \\ \hline
     C6: Control
     (Western World)&   government, shabbos, biden, occupational, enslavement, american, goy, fake &zog/zionist, shabbos goy, ashkenazi jewish, enslavement agenda, mind control, terrorist cabinet, kosher sandwich	   \\ \hline
     C7: Dystopia &   slaughtering, zionist, partner, rock, agent, stone, time, information, domain&online domain, zionist reptilian, zionist partners, black stone, latin brown, black rocks	   \\ \hline

      C8: ZOG &  Jewish, communist, zionist, zog, capitalist, like & ZOG, scott greer, elephant room, capitalist subhuman, machine Georgia, bloodline conglomeration, diana spencer, gay jewish \\ \hline
      C9: Jewish Mafia  &Jewish, rothschild, world, f**k, people, need, soros & jacob rothschild, khazarian mafia, klaus schwab, mafia minion, george soros, trillionaire jacob \\ \hline
     
    \end{tabular}
    \end{adjustbox}
    
    \label{tab:terms_both_uni_bigram}
\end{table*}
We now present a qualitative analysis of the results obtained by our method and reported in Tables~\ref{tab:terms_both_uni_bigram} and~\ref{tab:examples_concepts}. 
% 
% and~\ref{tab:examples_concepts}. %We labeled concepts C1 to C7 as follows. The labels are abbreviated in Table~\ref{tab:terms_both_uni_bigram}.
\begin{comment}
\begin{description}
\item[C1:] Accusation of Economic Control
\item[C2:] The Goyim Know (Accusation of Economic/Political Control)
\item[C3:] Accusation of Cultural Control
\item[C4:] Accusation of Political Control
\item[C5:] Accusation of Rejection of Christianity 
\item[C6:] Accusation of Control of western States (ZOG Conspiracy)
\item[C7:] Dystopian Antisemitism (Reptilian Conspiracy)   
\end{description}
\end{comment}
The tables show samples of prototypical terminology and posts per extracted concept. The results of Table~\ref{tab:terms_both_uni_bigram} were obtained using aspects\footnote{\cite{Kikkisetti2024UsingLT} focuses on coded terminology. This work doesn't so we disabled the filtering of non-coded terminology. We also return unigrams in addition to bi- and tri- grams.} of the term extraction methodology described in \cite{Kikkisetti2024UsingLT} on each of the concepts extracted by our method. Due to space limitations, Table~\ref{tab:examples_concepts} can be found in the Appendix. 
We conducted a qualitative analysis of these tables as follows. The majority of posts in C1 is about Jews controlling the world. Indeed, we frequently observe the term \texttt{New world order} which, according to the AJC Translate Hate Glossary (AJC) conveys, in its antisemitic use, the belief that Jews have created a power structure that they fully control at the economic, media and political levels.  Some frequent unigrams appearing in C1 are \texttt{evil, need, state, zionist, globalists, economic} while bigrams and trigrams include \texttt{new world order, world economic forum, deep state, banking cabal, state cabal}. As discussed in AJC, the term ``cabal" refers to a powerful group of (Jewish) individuals whose goal is to establish control while ``globalists", which also introduces the idea of dual loyalty, can refer to a Jewish elite bent on destroying the western world order. % We see these ideas expressed in the posts of table~\ref{tab:examples_concepts}. 
Based on the terms extracted for C1 (Table~\ref{tab:terms_both_uni_bigram}) and on the typical posts belonging to this category (Table~\ref{tab:examples_concepts}), we concluded that while the concept presents a mixture of control types, the dominant one was economic, and we thus label it \textbf{C1: accusation of economic control}. % and summarize them in the root node of the left tree in Figure~\ref{tab:tropes_tree}.  %Uni-gram and Bi-gram reflect Trope -- Jews have too much power and fear of Jews dominance see Figure . 
In C2, our approach grouped posts referring to the meme \texttt{Shut it down; The goyim know} into a single concept, which we labeled \textbf{C2: The goyim know}. This meme portrays, in a satirical way, Jews panicking because their ``secret plot to control the world" %including the media, economy and governments (AJC) 
has been discovered (AJC). Our algorithm's ss-structure shows that C2 was derived from C1 along with C3 and C4 described next. %{\color{red} I think that the tree of Figure 5 should include the cluster numbers. I don't see which one, C2 belongs to...  With the explanation, it makes sense for it to be isolated. Perhaps the node "Fear of Jewish control of the media should be turned back to Shut if down the Goyim know. There is nothing specific to control of the media in this data set, I think } %concept that the Jews basically control everything, such as the media, governments, and banks. The idea that everything upside down in the world is for the benefit of the Jews suggests that Jews have significant authority and play a significant role in global events using media and governments, which basically reflects the trope Jews power and fear of Jews dominance, especially in Media.  
In C3, we observe the terms \texttt{cultural, marxism, frankfurt, school} as well as \texttt{frankfurt school, white reproduction, political correctness}, etc. According to a report by the Southern Poverty Law Center (SPLC), these expressions refer to an antisemitic conspiracy theory that suggests that Cultural Marxism rooted in the Frankfurt School, a group of self-interested Jews, is responsible for progressive causes such as LGPTQ+, identity politics, and political correctness which are meant to destroy American values.\footnote{https://www.splcenter.org/fighting-hate/intelligence-report/2003/cultural-marxism-catching} We labeled that concept \textbf{C3: accusation of cultural control}. %In Figure 5, we labeled this node as Fear of Jewish Political control.
C4 is mostly centered around wars and historical events. The themes revolve around history including USSR, Russia, Ukraine, Britain, WWII and the current war between Russia and Ukraine. Holocaust denial is also present. We labeled this concept \textbf{C4: accusation of political control}. Two seemingly unrelated expressions appear in C4 including Mel Gibson and Moon Landing. Our Web investigation revealed that Mel Gibson, upon receiving probation for DUI in 2006, remarked to the policeman arresting him %whether he was Jewish and remarked 
that ``the Jews are responsible for all the wars in the world.''\footnote{\url{https://shorturl.at/dqyDH}}. %We assume that 
This comment (about wars) is what must have landed him into C4. Moon landing refers to the conspiracy theory that claims that the 1969 moon landing never happened (the footage was, supposedly, fake).\footnote{\url{https://en.wikipedia.org/wiki/Moon_landing_conspiracy_theories}} Moon landing landed in C4 most probably because of its role in the Cold War. Though the conspiracy theories revolving around it are not antisemitic in nature, it is worth noting that many proponents of these theories also support antisemitic ones and that some cross-overs sometimes take place.% For example, a French author recently interviewed on Iranian TV claims that Stanley Kubrick, the Jewish film-maker, was hired by NASA to make the fake footage in exchange for their funding of the making of "2001: A Space Odissey".
\footnote{E.g., https://www.memri.org/tv/french-author-guyenot-moon-landing-hoax-proves-america-is-empire-of-lies}  
%also exists for example, a coded term \texttt{moon landing} reflects many antisemitic remarks \footnote{\url{https://shorturl.at/pzBPR}}),  For example, a coded term which we find in the concept \texttt{mel gibson}, Mel received three years' probation remarking that ``the Jews are responsible for all the wars in the world.''\footnote{\url{https://shorturl.at/dqyDH}}. 
In C5, we observe religious themes represented by terms such as \texttt{Jesus, Jews bible, Jews commit, deicide, rabbinical tradition}. These terms and the text they center around them fit in with the historical antisemitic trope \texttt{Jews Killed Jesus Christ}. They are mixed with other terms such as \texttt{reptilian, Zionist reptilians} which refer to the reptilian conspiracy theory that suggests that reptilian aliens take the form of politicians in order to control the world.\footnote{\url{https://en.wikipedia.org/wiki/Reptilian_conspiracy_theory}} This, of course, gets back to the trope claiming that Jews control the world and is clearly antisemitic.\footnote{\url{https://shorturl.at/cfvEQ}} We labeled this concept \textbf{C5: rejection of Christianity} as we felt that despite the spread of this concept, that was the most representative theme. Next, in C6, the theme seems to revolve around the Zionist Occupation Government conspiracy theory that claims, once again, that Jews control the government of Western States.\footnote{\url{https://en.wikipedia.org/wiki/Zionist_Occupation_Government_conspiracy_theory}} 
%we find terms such as \texttt{fake jew, enslavement, ashkenazi jew}. These seem to represent themes of the Black Hebrew Israelites, an antisemitic group that claims that white Jews are not the real Jews, but that instead, they are.\footnote{\url{https://www.ajc.org/news/who-are-the-black-hebrew-israelites}}. 
This concept, which we labeled \textbf{C6: accusation of control of the Western States} is shown to be derived from C4 from the algorithm's ss-structure. Finally, C7 directly advocates for the slaughter of ``Zionists" using language rife in conspiratorial metaphors about, once again, reptilians, but other dystopian types of themes such as \texttt{online domains, time, information}. We labeled this concept \textbf{C7: Dystopian Antisemitism}.  C8, which we labeled \textbf{C8:Zionist Occupation Government (ZOG)} is derived from C6 (itself derived from C4). Finally, C9, labeled \textbf{C9: Jewish Mafia} lists names of Jewish millionaires or purported Jewish millionaires (e.g., Klaus Schwab) and is directly related to C8.% in the h-structure.
%We showed a few uni-gram and bi-gram in Table \ref{tab:trending_unigrams} and \ref{tab:trending_bigrasubms}. However, %In each trope (concept) from C1--C9, we observe unique combinations of antisemitic uni-gram and bi-gram, representing antisemitic concepts (trope).

\subsection{Quantitative Analysis (cont'd)} \label{sec:detailed-results}
\begin{figure*}[htp]
\centering
\subfloat[Affinity Propagation]{\includegraphics[width=0.33\textwidth]{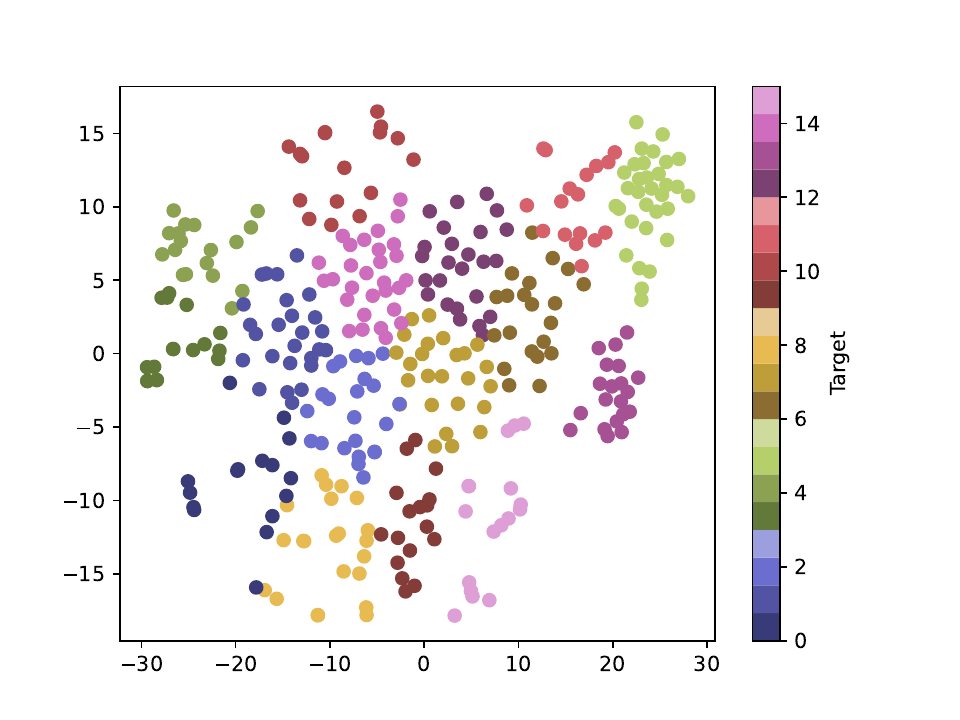}}%
    \subfloat[Birch]{\includegraphics[width=0.33\textwidth]{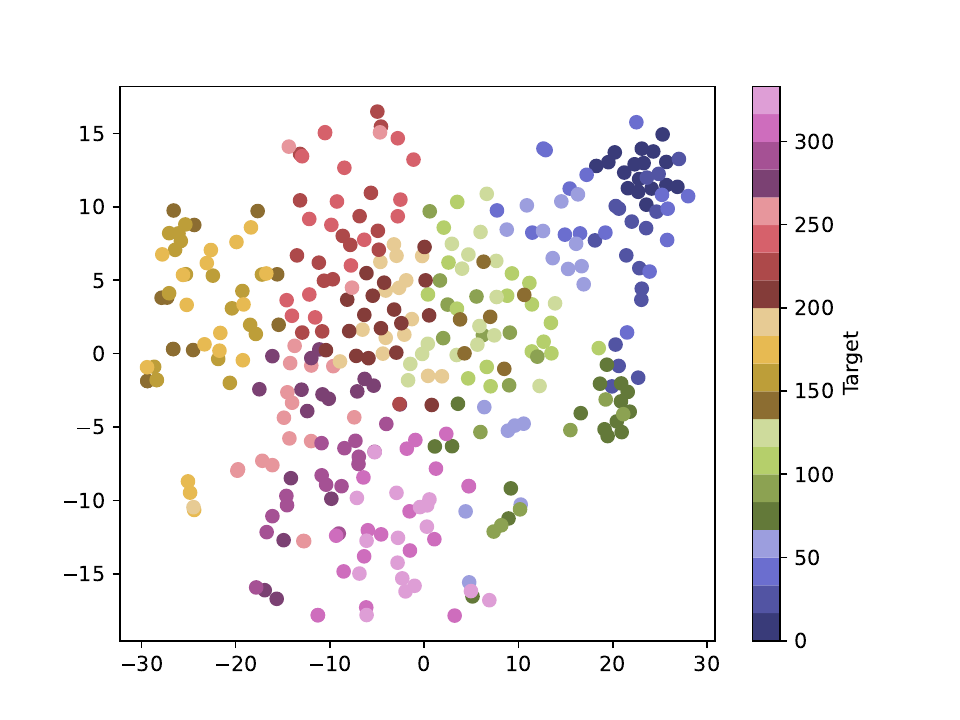}}%
    \subfloat[Birch (with K)]{\includegraphics[width=0.33\textwidth]{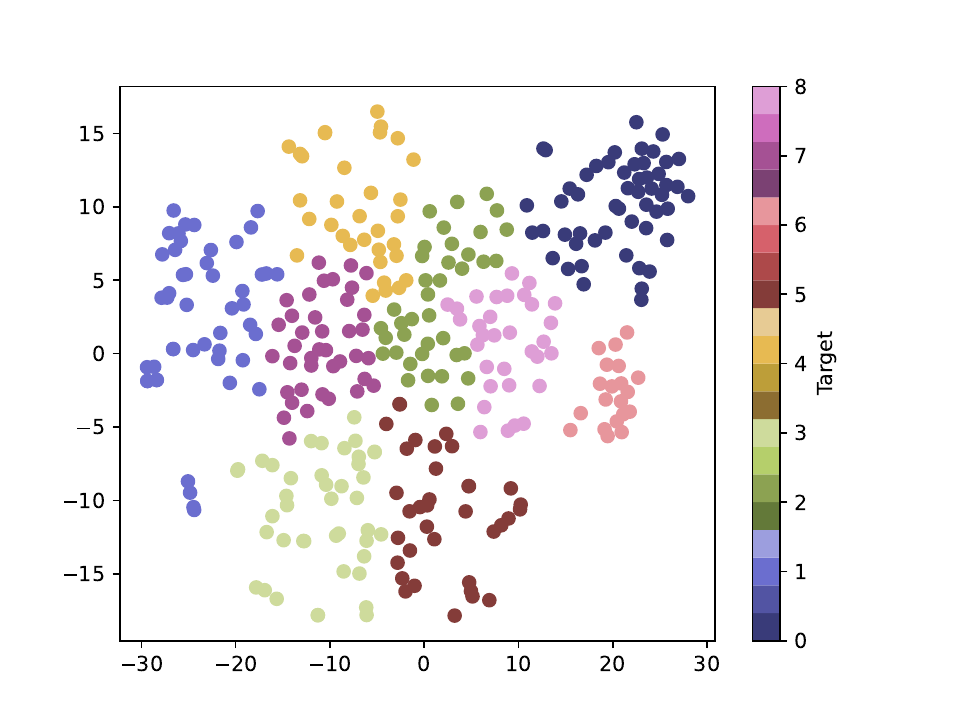}}
    
    \subfloat[Spectral]{\includegraphics[width=0.33\textwidth]{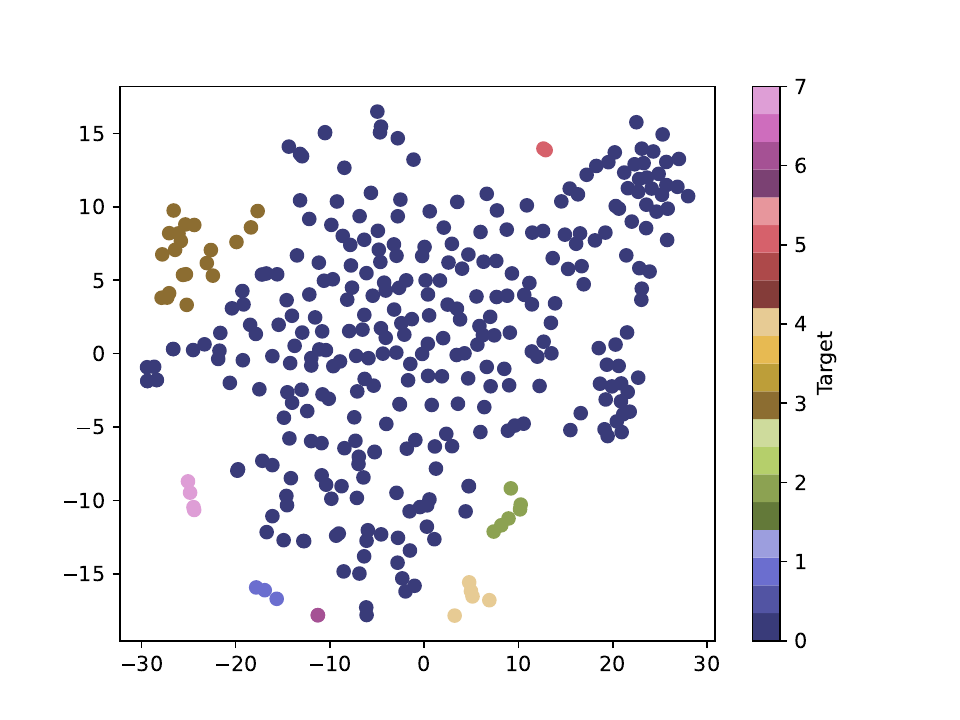}}%
    \subfloat[Gaussian]{\includegraphics[width=0.33\textwidth]{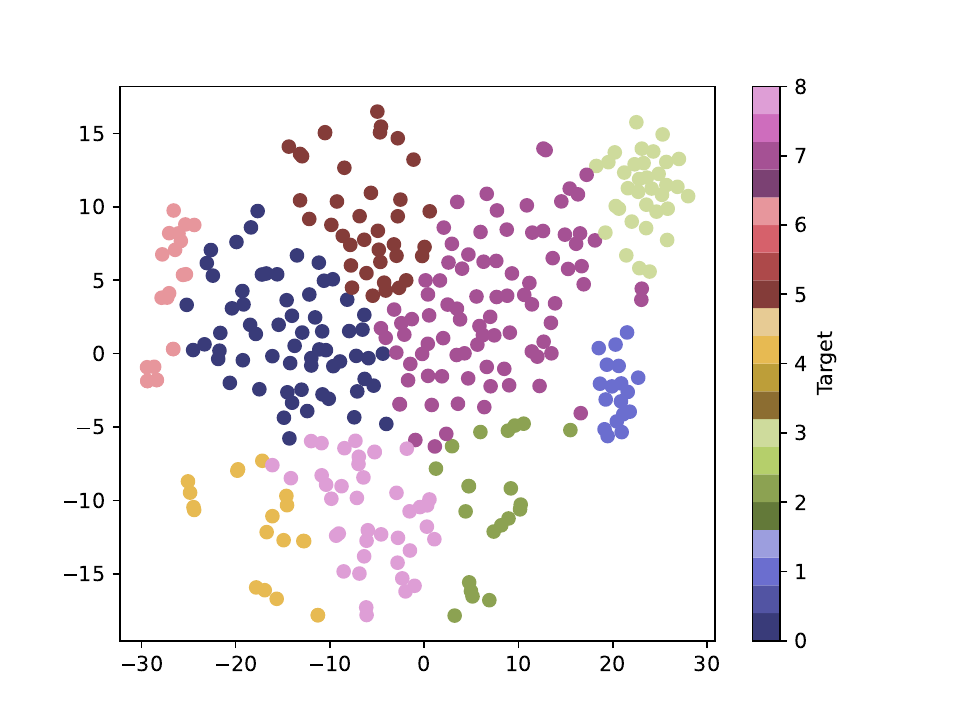}}%
    \subfloat[MeanShift]{\includegraphics[width=0.33\textwidth]{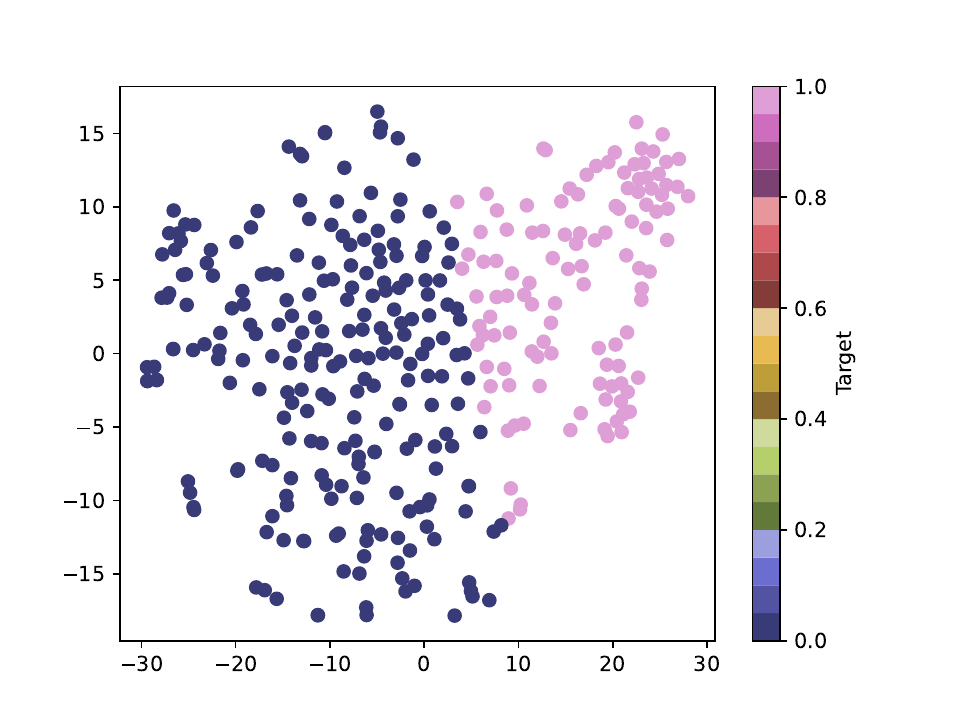}}\\
        \subfloat[Our Approach]{\includegraphics[width=0.33\textwidth]{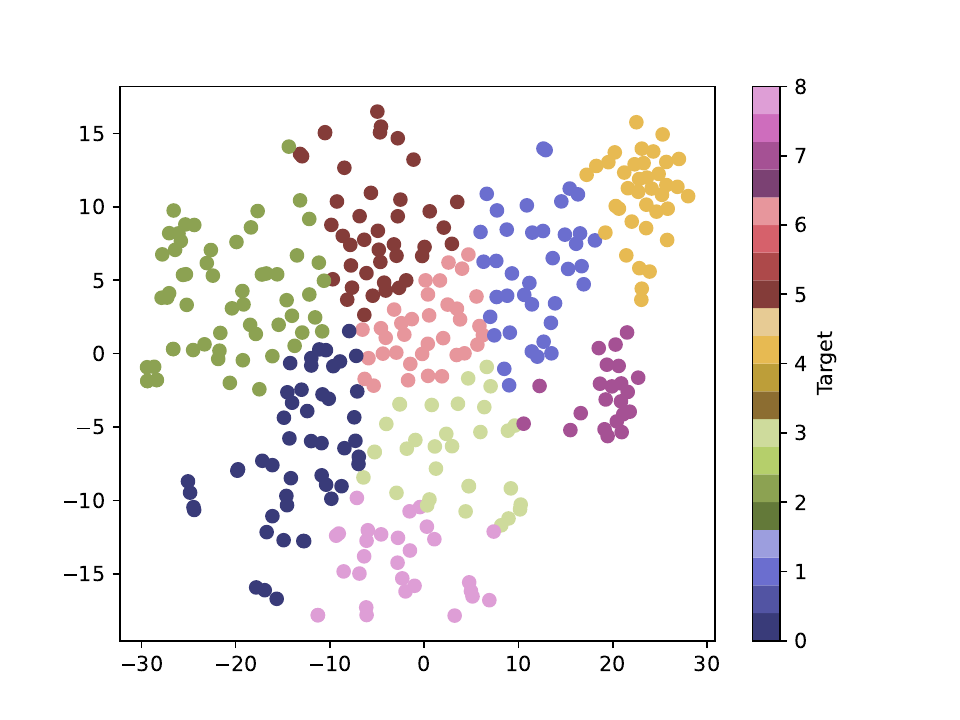}}

\caption{T-SNE projections of the clustering obtained by our approach and baseline clustering methods}
\label{tab:tsne}
\end{figure*}

Having described the concepts derived by our approach, we can now get back to the last two columns of Table~\ref{tab:baseline_results} that present the coverage by each approach of concepts C2 and C7 representing 15.90\% and {5\%} of the entire data set, respectively. C2 and C7 were chosen for a detailed analysis because they are the most compact and unambiguous concepts in the entire set of posts. We discuss their treatment by each of the clustering methods discussed in Section~\ref{sec:quant-an}. %We now summarize our observations. 
 Using our own approach described in Algorithm~\ref{alg:two}, coverage of concepts  C2 and C7 are 75.5\% %of the concept 
 and 100\%, respectively. We note that for C7, three other methods obtain 100\% coverage: Affinity, Birch, and Gaussian. Of these, Affinity is the only one that determines $k$, the number of clusters, on its own. The other two methods: Spectral and MeanShift receive the results ``Not Applicable" because, as seen in Figure~\ref{tab:tsne}, MeanShift learns only two concepts and, therefore, does not discriminate enough to learn small concepts and, in the case of Spectral, even though small concepts are learned, %in this case %$k$ was set to 9 prior to learning the algorithm, Figure~\ref{tab:tsne} show that Spectral's clustering is dominated by one huge cluster and eight very small outlying clusters. In more detail, using spectral clustering, 
 %we observed 
 our analysis revealed that C7 was merged with C4, with no clear separation between them. 
 C2 is more difficult to cover as seen in Table~\ref{tab:baseline_results}. Affinity splits C2 into two clusters covering 30.50\% and 23.72\%, respectively, of C2 or 54.22\% when considered together. When Birch is run with $k = 9$, it does pretty well, covering 67.69\% of C2. That is less than the 75.5\% obtained by our method, but it is close. Gaussian, on the other hand, only covers 33.89\% of C2. Once again, Spectral received the "Not Applicable" mention because C2 was distributed into more than one concept. Conversely, C2 was one of many amalgamated concepts in MeanShift. This analysis confirms that Birch works quite well on this domain, but that, based on this detailed analysis as well as the previously mentioned strength of our approach, it is preferable given our practical goals.

\section{Conclusion} \label{tab:conclusion} 
This work proposes a completely unsupervised machine learning method for monitoring the evolution of antisemitic discourse in a continual manner. As new posts arrive, they are presented to the system in a mini-batch fashion. Concepts representing new themes discussed in the posts are formed and added to the existing knowledge base while existing concepts are updated in accordance with the new information contained in the posts. Our quantitative and qualitative analyses reveal that our approach accurately describes the themes discussed in the posts, thus providing a useful monitoring tool for social media content. Our future work will add lifelong learning features to ensure that our method is robust enough to sustain long periods of online use. Our future work will also involve working in tandem with social scientists to assess how useful our tool is in answering their research questions. This will allow us to refine our method to make it more focused on its users' needs.
%Our future work also includes analyzing text from extremist platforms after the October 7 incident and discovering new emerging themes. Now, as there has been a significant increase in openly antisemitic posts after this incident, we also want to see if the antisemitic posts and terms are being made in coded language or not? If it is coded, what is the new addition, and if it is extremely harmful content, what is its hierarchy? 
In addition, we will add information visualization features to make the approach more user-friendly. Finally, we will apply our work to other kinds of hatred in social media.

 % \newpage

%Antisemitism is extremely provocative. Today, anonymous users or groups are exploiting social media to spread antisemitic posts, which causes hate crimes and conspiracy theories. In order to combat this, this paper proposes an approach for discovering antisemitic tropes from the extremist dataset. We also propose an approach for extracting antisemitic terminology from the tropes. Our results are useful in distinguishing between different types and layers of communication on extremist platforms. These results also provide contextual analysis of tropes and antisemitic terms to understand ``in what context term or tropes used''.  We found that some terms have become so common that they appear in more than one concept, which makes it clear that a new concept does not necessarily encompass the new antisemitic terms, however a new theme can be shaped by using these terms. (in-progress)

\bibliographystyle{ieeetr}
\textbf{\bibliography{ref}}

% \newpage

\section{Appendix}
% \begin{figure}[ht]
%     \begin{minipage}{.33\textwidth}
%         \centering
%         \includegraphics[width=\linewidth]{cls/clusters/affinity.pdf}
%        \subcaption{Affinity Propagation}
%     \end{minipage}%
%     \begin{minipage}{.33\textwidth}
%         \centering
%         \includegraphics[width=\linewidth]{cls/clusters/birch_no_clusters.pdf}
% \subcaption{Birch}
%     \end{minipage}%
%     \begin{minipage}{.33\textwidth}
%         \centering
%         \includegraphics[width=\linewidth]{cls/clusters/birch_with_clusters.pdf}
% \subcaption{Birch (with K)}
%     \end{minipage}

%      \begin{minipage}{.33\textwidth}
%         \centering
%         \includegraphics[width=\linewidth]{cls/clusters/spectral.pdf}
%        \subcaption{Spectral}
%     \end{minipage}%
%     \begin{minipage}{.33\textwidth}
%         \centering
%         \includegraphics[width=\linewidth]{cls/clusters/gm.pdf}
% \subcaption{Gaussian}
%     \end{minipage}%
%     \begin{minipage}{.33\textwidth}
%         \centering
%         \includegraphics[width=\linewidth]{cls/clusters/meanshift.pdf}
% \subcaption{MeanShift}
%     \end{minipage}
%      \begin{center}
%     \begin{minipage}{.33\textwidth}
%         \centering
%         \includegraphics[width=\linewidth]{cls/clusters/ours.pdf}
%         \subcaption{Our Approach}
%     \end{minipage}
%     \end{center}
%     \begin{minipage}{.99\textwidth}
%     \captionsetup{justification=centering}
%     \caption{T-SNE projections of the clustering obtained by our approach and baseline clustering methods}
    
%     \label{tab:tsne}
%     \end{minipage}
% \end{figure}

\begin{table*}
  \centering
  \caption{Examples from our data set of posts found on extreme social platforms along with the concept they belong to and the terminology they use as extracted by the approach in \cite{Kikkisetti2024UsingLT}}
  \begin{tabular}{| m{16cm} | m{4em} |}
    \hline
    \textbf{Posts} & \textbf{Concepts} \\
    \hline
    these poor souls were murdered by the {\color{red}cabal} {\color{red}world economic forum} evil mad scientists elites. they should all be in front of a patriot firing squad & C1 \\\hline
    the illegal us corp of american government must be eradicated closed permanently all the {\color{red}cabal} ,{\color{red}illuminate}, {\color{red}world economic forum}, world health organization, united nation, north atlantic treaty organization, {\color{red}new world order} and its sub cults all must be arrested and fully prosecuted world wide door to door & C1\\\hline
    % this evil needs to put on trial with the world economic forum \textbf{globalists} \textbf{cabal} enemies of humanity & C1 \\\hline
    the {\color{red}goyim know} about isis. & C2  \\
    \hline
   {\color{red}shut it down, the goyim know} & C2  \\\hline
    % me at my last klan rally: i exposed the \textbf{kike} and it made the \textbf{goyim know} woo hoo! & C2  \\
    % \hline
    advertising should be banned. it has been used as a social manipulation tool for decades. just watch mad men. it started with the {\color{red}frankfurt school}. the birth of {\color{red}cultural marxism}. how the {\color{red}frankfurt school} changed america & C3 \\\hline
    we are in a war with {\color{red}frankfurt school} {\color{red}cultural marxism} . harris is an apostle of that woke cult. please stop conflating liberalism with left wing marxism when we need to divide them starting with the difficult task of convincing boomer liberals they have been fooled since teenage hood.& C3 \\\hline
    % that would not happen in a fascist society. what you are witnessing in the west is trotskyst woke \textbf{cultural marxism} coupled with capitalist neoliberism. 	& C3 \\\hline
% 
     you also need to provide all {\color{red}zionist agents} embedded in the media organizations, zionist codenames, zionist handlers, zionist caves they report to, zionist time travellers giving them information, space time information about zionists, {\color{red}zionist "partner/s"}, zionist alliances, zionist sponsored businesses, zionist media organizations, {\color{red}zionist} agents embedded in the national security apparatuses,{\color{red} zionist agents}' names censoring news and content, etc., etc. ... & C4 \\\hline
     the united states government is controlled by {\color{red}interest groups} that are only seeking to enlarge their own power. the us government does not represent the will of the citizenry, and condemning it is not a condemnation on the principles of freedom, democracy, etc.the usa is being set up to fail.the rootless {\color{red}cosmopolitan elite} have been constructing elaborate safehouses for decades in preparation for this... & C4 \\ \hline
     % the jews had been trying for decades to instigate civil war they succeeded once according to even the most jewish of history books, we all know both jewish communist revolutions happened even if they ultimately failed as they killed millions. not even the kikes will debate this, they just say it wasn't important, don't put it in the textbooks and so on and almost staged a successful communist revolutionthere you go!zyklon bi love how even after the war americans continued delousing migrant workers from all over the world - often via direct application - and these migrants did not die. but when a jew gets so much as a photon reflected from the letter z into its eye, he becomes a pile of ashes which can never be photographed or weighed, such is the totality of his destruction & C4 \\\hline
     did jesus came to his own? yes but his own did not receive him? that is correct. so jews killed jesus? yes. so jesus was jewish? no, jews are not the jews of the bible so how did jews commit deicide if they are not the jews from the bible? hurr durr & C5 \\\hline
     he was but the definition of jew changed entirely after the fulfilment of the prophecy those who calls themselves jews now are not the real jews they are the {\color{red}rabbinical tradition} that rejected christ & C5 \\\hline
     % he created his universe by three forms of expression numbers letters and words=826 the first one hundred and forty four decimal digits of pi sum to six sixty six=826 american news events have been nothing but a rothschild crafted government stage show=826 jesus as well as his wife mary magdalene teaches the path of wholeness toward holiness=826 spiritually oppressing my people is cause for their souls to be destroyed=826 & C5 \\\hline

    meet the new shabbos, same as the old shabbos stop eating the {\color{red}kosher sandwich} goy {\color{red}zionist occupied government}.& C6 \\\hline
     {\color{red}zog}, {\color{red}zionist occupied government}. secretary of state=zog. attorney general=zog. senate majority leader=zog. head of dhs=zog 98 percent of americans are non jews.& C6 \\\hline
     % zog = zionist occupied government. zoc = zionist occupied christianity. fake preachers are worse than fake jews. 	 & C6 \\\hline
     are the jews of belgium, providing the zionist codenames, zionist agent names, names of the {\color{red}zionist reptilians}, names of the zionist "partner/s", zionist caves, space time information about zionist, etc need full disclosure from the zionists, {\color{red}zionist reptilians}, zionist agents, zionist "partner/s", and reptilian "partner/s", etc. afterwards, i will slaughter them all, including their women, and their children. i will spare no one.but then again, i prefer genocides, mass erasures, etc., rather than patches.& C7 \\\hline
     i also need the names of all zionist partner/s, all zionist agents, zionist caves, zionist time travel agents, zionist time portal technology, zionist advanced technology, and {\color{red}zionist reptilian} "partner/s".and all {\color{red}zionist reptilians} embedded in albania, and zionist reptilian "partner/s" embedded in albania.all can be published in the {\color{red}online domain}, on the internet, available for all to dissect such information.afterwards, i will slaughter them all. including zionist women, and zionist children, zionist "partner/s", and those that interfere on behalf of zionists. 	 & C7 \\\hline
     % how are the zionist purges going?are the zionists of the nsa providing information on other zionists, on zionist reptilians, their codenames, their caves, zionist reptilian alliances, space time information about zionists, zionist "partner/s", zionist advanced technology, and more specifically zionist time portal and time travel, etc.all such information should be disclosed, published etc.afterwards, slaughter all zionists, including zionist "partner/s", and those that interfere on behalf of zionists. spare no one.and install blood type b's, and oversight by those aligned with the wellbeing and peaceful coexistence, of the blood type b's, first, and then a's. & C7 \\\hline
     you can post about the gae all you want and not get banned on any platform, like {\color{red}scott greer} for example. start tweeting about the {\color{red}zionist occupied government} and expect to be censored and banned. lesson there. & C8 \\\hline
     you got a {\color{red}gay jewish} communist takeover that is as communist as the chinese, without putting communist in the name.enjoy being spied, subverted, and psyoped.&C8\\\hline
     {\color{red}klaus schwab} is a {\color{red}rothschild} on his mother side. he answers to the trillionaire, {\color{red}jacob rothschild} and his {\color{red}khazarian mafia minions} & C9 \\\hline
    radical soros backed group boosts warnock campaign funding ahead of run off billionaire radical {\color{red}george soros} has worked his entire life to control the democrat party and destroy the united states. he has a deep seated hatred of the west and promotes its collapse, pouring millions of dollars into campaigns and causes that further his radical agenda. & C9 \\\hline
  \end{tabular}

  \label{tab:examples_concepts}
\end{table*}


\end{document}